\documentclass{article}
\usepackage{ltwol2e2}

\usepackage{psfig}

\def\be{\begin{equation}}
\def\ee{\end{equation}}
\def\bea{\begin{eqnarray}}
\def\eea{\end{eqnarray}}
\def\VEV{\langle H \rangle}
\def\VEVi{\langle H_1 \rangle}
\def\VEVii{\langle H_2 \rangle}
\def\sfrac#1#2{{\textstyle\frac#1#2}}

\begin{document}

\title{PREDICTIONS FOR SUSY PARTICLE MASSES FROM ELECTROWEAK BARYOGENESIS}

\author{J.~M.~CLINE}

\address{McGill University, Dept.~of Physics, 3600 University St.,
Montr\'eal (Qc) H3A 2T8, Canada\\E-mail: jcline@physics.mcgill.ca}   


\twocolumn[\maketitle\abstracts{In collaboration with
 G.D.~Moore,\cite{cm} the electroweak phase transition in
the minimal supersymmetric standard model is studied using the two-loop
effective potential.  We make a comprehensive search of the MSSM
parameter space consistent with electroweak baryogenesis, taking into
account various factors:  the latest experimental constraints on the
Higgs boson mass and the $\rho$ parameter, the possibility of
significant squark and Higgs boson mixing, and the exact rate of bubble
nucleation and sphaleron transitions.  Most of the baryogenesis-allowed
regions of parameter space will be probed by LEP 200, hence the Higgs
boson is likely to be discovered soon if the baryon asymmetry was
indeed created during the electroweak phase transition.}]

\section{Introduction}
An attractive possibility for explaining the baryon asymmetry
of the universe,
\be
	{n_B - n_{\bar B} \over n_\gamma} \sim 10^{-10}
\ee
is electroweak baryogenesis, combined with supersymmetry.  This is one
of the only pictures which makes fairly definite low-energy
predictions, which are in effect being tested now at LEP and the
Tevatron by their searches for the Higgs boson and the top 
squark.\cite{cm,cqw2}

Let us review the basic picture of electroweak baryogenesis.  It
assumes that the electroweak phase transition (EWPT) is strongly first
order, hence bubbles of the true vacuum, with nonzero Higgs VEV $\VEV$,
nucleate inside the false vacuum, at a critical temperature $T_c$ near
100 GeV (fig.~1).  Outside the bubble, anomalous baryon-violating
interactions (sphalerons), present in the Standard Model, are occuring
much faster than the Hubble expansion rate.  Inside the bubble, if
$\VEV/T_c$ is larger than $\sim 1$, these interactions are out of
equilibrium.  In addition there must be CP-violating interactions at
the bubble wall, which cause an asymmetry in the reflection probability
for particles versus antiparticles (and left-handed versus right-handed
particles) from the wall.  This causes a build-up of CP asymmetry
in front of the wall, which the sphalerons attempt to erase.  But
in so doing, they create a baryon asymmetry in front of the wall,
which gradually falls behind the wall due to the steady expansion
of the latter into the plasma, and collisions of the reflected
particles with other particles in the plasma.  This baryon excess
survives to become the baryonic matter of the present-day universe.

\begin{figure}
\centerline{\psfig{figure=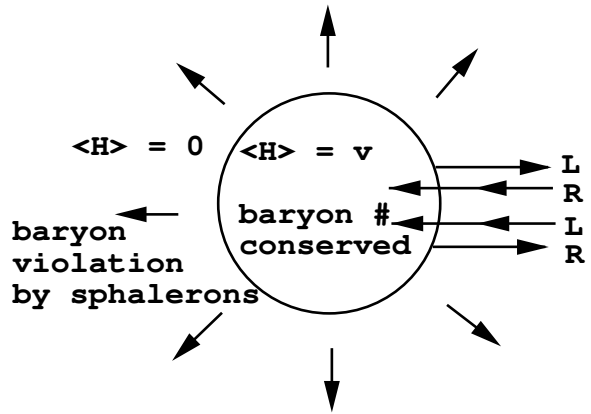,width=3.0in}}
\caption{Expanding bubble during the electroweak phase
transition.  Left-handed fermions reflect into right-handed ones
and vice versa, but with different probabilities.}
\end{figure}

There have been attempts to implement electroweak baryogenesis in other
ways, for example, using cosmic strings instead of bubbles.  The idea
is that $\VEV$ can be suppressed inside the strings; then strings could
work similarly to bubbles except that the baryon ($B$) violation is
going on inside of the strings as they sweep through space, while $B$
is conserved outside because the temperature is assumed to be much less
than $T_c$.  We have made a quantitative study of all aspects of this
proposal \cite{cmer} (except for the possibility of superconducting
strings which carry an enormous current), and concluded that it falls
short of being able to produce the required $B$ asymmetry by 10 orders
of magnitude.  One reason for the failure is that the CP violation
scales with the string velocity ($v$) squared, whereas the density of a
string network scales like $v^{-2}$.  It is therefore impossible to
tune the density to enhance the asymmetry.  Another reason for the
failure is that sphalerons are typically large compared to strings, and
their energy, $E_{sph}$, is increased if they are squashed so as to fit
inside a string.  This suppresses their likelihood, hence the rate of
$B$ violation, by a Boltzmann factor, $e^{-E_{sph}/T}$.

Thus the expanding bubble picture is the most likely realization of 
electroweak baryogenesis.  And it is also highly constrained, since
it is not easy to get a large enough asymmetry.  One of the major
challenges is in getting the phase transition to be strong enough
so that $\VEV/T_c \ge 1$.

\section{Strength of the Phase Transition}

In the Standard Model, the EWPT is second order, as illustrated in
figure 2a.  The effective potential, $V(H)$, never develops a barrier
between the high-temperature, symmetric phase minimum ($H=0$) and the
low-$T$, broken phase one ($H\neq 0$).  Bubbles do not form in this
case, so baryogenesis as outlined above cannot occur.  However, the
EWPT can be first order in the Minimal Supersymmetric Standard Model
(MSSM) if one of the top squarks (the mostly right-handed one) is
sufficiently light,\cite{cqw2,jre,bjls,lr1} giving an effective
potential of the form shown in figure 2b.

\begin{figure}
\centerline{\psfig{figure=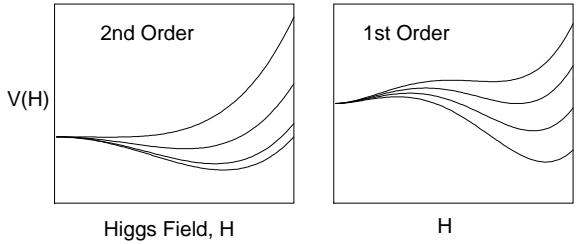,width=3.5in}}
\caption{(a) Higgs effective potential for several temperatures, with
 a second order transition. (b) Same, but with a first order transition.}
\end{figure}

In the presence of light stops (light compared to the temperature),
$V(H)$ gets finite-temperature contributions from vacuum loop diagrams
containing stops and possibly gluons, like those of figure 3.
The propagators are evaluated
at an arbitrary value of the background Higgs fields, of which there
are two in the MSSM.  The squark masses appearing in the propagators
are the Higgs-field-dependent eigenvalues of the mass matrix
\be
	{\cal M}_{\tilde t}^2 \cong
	\left(\begin{array}{cc} m^2_Q + y^2 H_2^2 + O(m^2_Z) &
	y(A_t\,H_2 - \mu\, H_1) \\ y(A_t\,H_2 - \mu\, H_1) &
	m^2_U + y^2 H_2^2 + O(m^2_Z)\\ \end{array}
	\right).
\label{massmatrix}
\ee
Because the top Yukawa coupling $y$ is large, the stops couple
strongly to the Higgs field $H_2$, and this is why they have a
potentially strong effect on $V(H)$.

\begin{figure}
\centerline{\psfig{figure=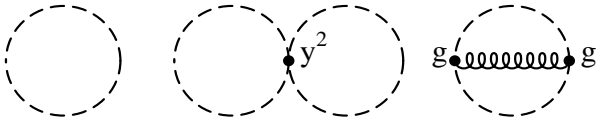,width=3.1in}}
\caption{The most important one- and two-loop stop diagrams contributing to
the effective potential.}
\end{figure}

We have computed $V(H)$ to two loops, including mixing effects 
between the stops and between the heavy and light Higgs fields.
We then searched the MSSM parameter space to find regions where
the phase transition is strong enough for successful baryogenesis,
while still compatible with other experimental constraints. 

\section{Important Parameters and Constraints}

Now I will discuss the parameters which have the strongest 
effect on the EWPT.

{\bf Lightest Higgs mass, ${\mathbf m_h}.$}  The transition is 
strongest if $m_h$ is small.  Moreover, the light Higgs field
$h$ is related to the flavor eigenstates $H_i$ by a mixing angle
$\alpha$,
\be
	h = \sin\alpha\, H_1 + \cos\alpha\, H_2,
\ee
and the phase transition is also strongest when $\sin^2(\alpha-\beta)
\approx 1$, where $\tan\beta = \VEVii/\VEVi$.  This is illustrated
in fig.~4a, which shows the distribution of allowed points from our
Monte Carlo of the parameter space, in the plane of $\sin^2(\alpha-\beta)$
and $m_h$.  The region to the left of the line is experimentally 
excluded by LEP.
\begin{figure}
\centerline{\psfig{figure=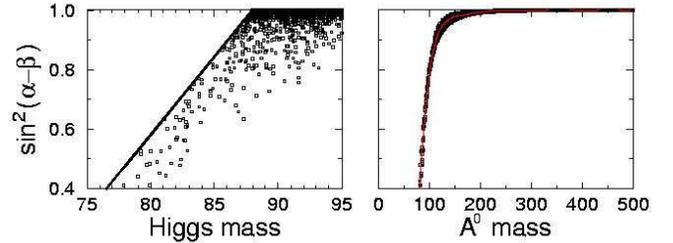,width=3.4in}}
\caption{(a) Scatter plot from Monte Carlo in plane of Higgs
mixing angles versus lightest Higgs mass (in GeV); 
(b) Same, but with the mass of the $A^0$ Higgs boson.}
\end{figure}

{\bf CP-odd Higgs mass, ${\mathbf m_{A^0}}.$} A strong phase transition
also favors the pseudoscalar Higgs boson, $A^0$, being heavy.  In fact,
this explains the preference  for $\sin^2(\alpha-\beta)\approx 1$,
because there is a strong correlation between $m_{A^0}$ and 
$\sin^2(\alpha-\beta)$, as shown in fig.~4b.  In the limit where both
become large, the Higgs sector of the MSSM becomes SM-like, with all
the heavy Higgs bosons decoupling.

{\bf tan}${\mathbf\beta}$ {\bf and} ${\mathbf m^2_U}$. We find a lower limit
on the value of $\tan\beta$ which gives a strong enough transition:
\be
	\tan\beta > 2.1
\ee
This is coming largely from the fact that $m_h$ depends on $\tan\beta$,
\bea
	m^2_h &=& \sfrac12\left[m^2_A+m^2_Z-\sqrt{(m^2_A+m^2_Z)^2
	-4m^2_Z m^2_A \cos^2 2\beta}\right]\nonumber\\ 
	&+& O\left[(m_t^4/v^2)\ln\left(m_{\tilde t_1}m_{\tilde t_2}
	/m_t^2\right)\right],
\eea
in such a way that its tree-level value vanishes when $\tan\beta=1$.
Thus to satisfy the experimental constraints on the Higgs mass, one
must take $\tan\beta$ to be greater than some minimum value.  As for
larger values of $\tan\beta$, these tend to weaken the phase transition.
However this effect can be counteracted by enhancing the stop contribution
to $V(H)$, {\it i.e.,} by making the stop lighter.  To make the
right-handed stop sufficiently light, we must in fact take its
mass parameter $m^2_U$ to be negative!  This combines with the term
$m^2_t$ in eq.~(\ref{massmatrix}) to give an overall positive mass
squared, except before the EWPT when $m_t = 0$.  Figure 5 shows how 
the strength of the phase transition (measured as $\VEV/T$, where
$\VEV = \sqrt{\VEVi^2+\VEVii^2}$\ ) depends on $\tan\beta$ and on
$\tilde m_U\equiv m^2_U/|m_U|$.

\begin{figure}
\centerline{\psfig{figure=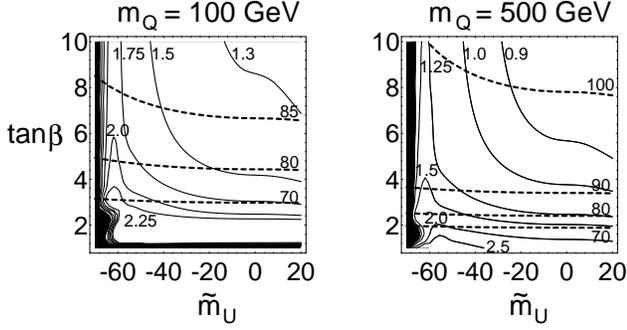,height=1.75in}}
\caption{Contours of $v/T$ (solid) and Higgs mass (dashed)
in the plane of $\tan\beta$ and $\tilde m_U\equiv m^2_U/|m_U|$ (in GeV), 
for $m_Q = 100$ and 500 GeV, respectively, at zero squark mixing
($\mu=A_t=0$).  The potential has color-breaking minima
in the black regions near $\tilde m_U = -70$ GeV.}
\end{figure}

{\bf Squark mixing parameters,} $\mathbf\mu$ {\bf and} $\mathbf A_t$.
Although the EWPT is generally stronger for small values of the
squark mixing parameters (appearing in the off-diagonal elements
of the mass matrix (\ref{massmatrix})), the preference is rather weak.
Histograms for $\mu$,  $A_t$ and the squark left-right mixing angle
$\theta$ are shown in figure 6.  
Small values of $|\mu|$ are experimentally excluded by chargino
and neutralino searches, explaining the absence of points near
$\mu = 0$.

\begin{figure}
\centerline{\psfig{figure=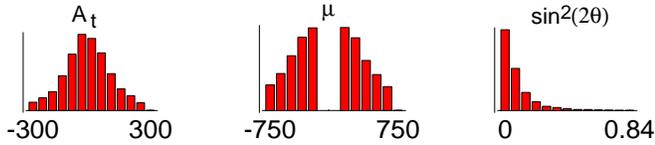,width=3.5in}}
\caption{Histograms for squark mixing parameters and mixing
angle from Monte Carlo.  Mass parameters are in GeV.}
\end{figure}

{\bf Left-handed stop mass.}  The left-handed stop mass parameter
is bounded from below,
\be
	m_Q > 130 {\rm\ GeV}.
\ee
This comes from the requirement that the contributions to the $\rho$
parameter from the squarks not exceed precision electroweak bounds
(we take $\Delta\rho < 1.5\times 10^{-3}$).  Because the light Higgs mass
depends on $m_Q$ through radiative corrections, it has a noticeable
effect on the baryogenesis limits on these parameters.  The maximum
allowed light Higgs mass, and the minimum allowed values of $\tan\beta$,
depend on $\hat m_Q\equiv (m_Q/100$ GeV) as
\bea
\label{fits}
	m_h &\le & 85.9 + 9.2 \ln(\hat m_Q) \ {\rm GeV}
	\nonumber\\
	\tan\beta &\ge & 
	(0.03 + 0.076\,\hat m_Q - 0.0031\,\hat m_Q^2)^{-1}.
\eea
The corresponding scatter plots from the Monte Carlo, 
for these quantities versus $m_Q$ are shown in figure 7.

\begin{figure}
\centerline{\psfig{figure=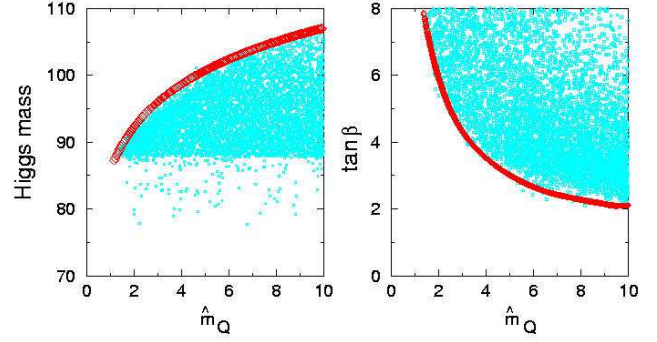,width=3.4in}}
\caption{Scatter plots from Monte Carlo for $m_h$ and $\tan\beta$
versus $\hat m_Q\equiv m_Q/(100$ GeV).}
\end{figure}

\section{Strong Phase Transition at Small $m_{A^0}$?}

We have seen that the majority of accepted points in our Monte Carlo
are those with large values of $m_{A^0}$.  There are some rare
exceptions, as can be seen in fig.\ 7a: the sparse points below $m_h =
87$ GeV are those with $m_{A^0}\sim 100$ GeV.  Although they
infrequently give a strong enough transition, they could be important
for the following reason.  Most of the contributions to the CP
asymmetry that gives baryogenesis are proportional to the amount by
which $\tan\beta$ changes inside the bubble wall, and this is strongly
correlated with the value of $m_{A^0}$.  

The concept of $\Delta\beta$, the deviation in $\beta$, is illustrated
in fig.\ 8.  The path in field space taken by the Higgs field as it
goes from inside the bubble ($\VEV > 0$) to outside ($\VEV = 0$) is
shown for the case of $m_{A^0} = \infty$, where it is a straight line,
and for $m_{A^0} \sim 100$ GeV, which gives some curvature to the
path.  The maximum angular deviation from straightness can be called
$\Delta\beta$, and computed from the effective potential.  (For
technical reasons we define $\Delta\beta\equiv {\rm max}_v[v
(\beta(v)-\beta(v_c))]/v_c $, where $v$ is the value of $\VEV$ at any
point inside the bubble wall.)  In figure 9 we show how $\Delta\beta$
\begin{figure}
\centerline{\psfig{figure=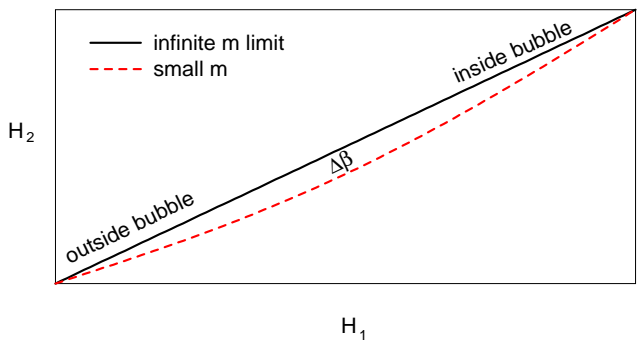,width=3.5in}}
\caption{Trajectories in Higgs field space for going through
the bubble wall, in the limits of large and small $m_{A^0}$.}
\end{figure}
is correlated with $m_{A^0}$ and the frequency of $\Delta\beta$
values.  Typically $\Delta\beta$ is quite small, $10^{-3}$, and only
very rarely reaches $10^{-2}$, as also found by others.\cite{mqs}

Since most electroweak
baryogenesis mechanisms have $n_B\propto \Delta\beta$, 
this gives an additional suppression in
the baryon asymmetry that can be produced in electroweak mechanisms,
which has been overlooked by some authors.  Interestingly,
there is one contribution to the CP asymmetry which has been shown to
be unsuppressed \cite{cjk} in the limit of vanishing $\Delta\beta$.
Charginos, which have the Dirac mass matrix
\be
	{\cal M}_{\tilde W\tilde h} = \left(\begin{array}{cc}
	m_2 & gH_2/\sqrt{2} \\ gH_1/\sqrt{2} & \mu  \end{array}
	\right),
\ee
can experience a CP-violating force while traversing the bubble
wall if the $\mu$ parameter is complex, even if $H_1/H_2$ remains
constant in the wall. 

\begin{figure}
\centerline{\psfig{figure=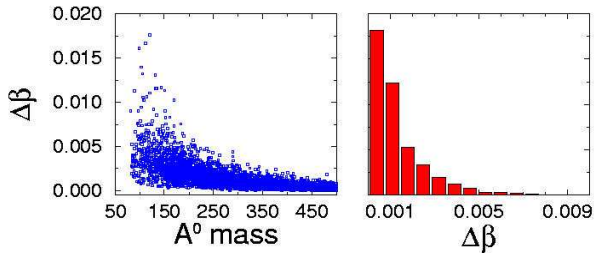,width=3.3in}}
\caption{(a) Maximum deviation in weighted Higgs VEV
orientation, $\Delta\beta\equiv {\rm max}_v[v
(\beta(v)-\beta(v_c))]/v_c $, inside bubble wall, as a function of
$m_{A^0}$; (b) distribution of $\Delta\beta$ values.}
\end{figure}

\section{Squark Masses and Mixing} 

Previous studies of the EWPT have emphasized the weakening effect that
squark mixing has on the phase transition strength.  We have already
pointed out that the Monte Carlo, although favoring small mixing
between the left- and right-handed stops, does not strongly exclude
large mixing: see figure 6 and eq.~(\ref{massmatrix}).  Since nonzero
values of $\mu$ are needed to get CP violation, this is fortunate!

As for the stop masses, the distributions of the relevant parameters
can be seen in figure 10.  The left-handed stop (which in the limit of
large $m_Q$ has mass approximately equal to $m_Q$) is usually much
heavier than the right-handed one, but can be as light as 116 GeV.  The
right-handed one is always relatively light, in the range 119 GeV $<
m_{\tilde t_1} < 172 $ GeV.  Such a light squark is potentially
discoverable at the Tevatron, but this depends strongly on the 
value of gluino
mass, $m_{\tilde g}$.  The discovery potential is greatly suppressed if
$m_{\tilde g} > 300$ GeV.

\begin{figure}
\centerline{\psfig{figure=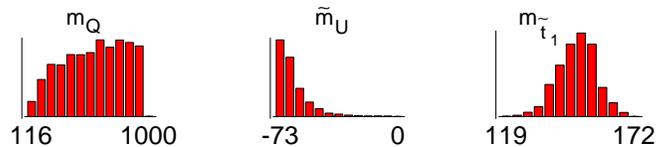,width=3.4in}}
\caption{Monte Carlo distributions for the mass  parameters
of (a) the left stop and (b) the right stop; and (c) the actual mass
of the right stop, in GeV.}
\end{figure}

\section{Sphaleron Rate versus Higgs VEV;\\ Latent Heat}

\begin{figure}
\centerline{\psfig{figure=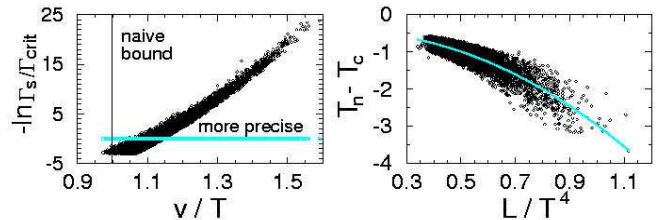,width=3.4in}}
\caption{(a) Correlation of the sphaleron rate with $v/T$;
below the line would be ruled out baryon preservation; 
(b) $\Delta T\equiv T_n - T_c$ in GeV versus
$\Lambda \equiv({\rm latent \; heat})/T^4$, fit by the function $\Delta
T = -0.5 + 0.5 \Lambda - 2.9 \Lambda^2$.}
\end{figure}

In this work we have also made a detailed study of the exact criterion
for a strong enough phase transition.  Recall that the requirement is
that sphaleron interactions inside the bubbles must be slow enough so
that the baryon number produced does not get erased afterwards.  The
naive condition for fulfilling this is that $\VEV/T_c\ge 1$, but the
more exact criterion is that the rate of sphaleron interactions,
$\Gamma_{s}$, must exceed some critical lower value, $\Gamma_{crit}$,
derived in ref.~[1].  By plotting $-\ln \Gamma_{s}/\Gamma_{crit}$
versus $v/T$ (fig.~11a), we can see how far off the naive condition
is.  The figure shows that a significant number of trial parameter sets
are excluded by the exact bound, although they may pass the $\VEV/T >1$
condition.  However, we have tried to correct for the fact that the
effective potential approach underestimates the strength of the phase
transition by about 10\%, compared to nonperturbative lattice
results.\cite{lr1}  Because of this, our threshold for acceptance
is somewhat looser than $\Gamma_{s}=\Gamma_{crit}$.

We have also studied the issue of reheating after the onset of the phase 
transition, which is related to the latent heat $L$.  $L$ is defined
to be the difference in $dV/d\ln T$ between the symmetric ($\VEV=0$)
and broken ($\VEV\neq 0$) phases.  Although the phase transition
becomes energetically possible starting at the critical temperature
$T_c$, where the two phases are degenerate in energy, stable bubbles
do not start to appear until the somewhat lower nucleation temperature,
$T_n$.  If the transition is strong enough, entropy production at
the bubble walls could conceivably reheat the universe all the way back to
$T_c$.  In general the reheat temperature ($T_r$) is given by
\bea 
{T_c - T_r\over T_c - T_n} &=& 1 - {30 L\over g_*\pi^2(T_c^4-T_n^4)}
\nonumber\\
	&\cong& 1 - {15 L \over 2 g_*\pi^2 T_c^{3}(T_c-T_n)},
\eea
which approaches zero if $T_r\to T_c$ (here $g_*$ is the number of
relativistic degrees of freedom in the plasma).  We find a correlation
between $L$ and $T_c-T_n$ (fig. 11b) so that the right hand side can be
thought of as being a function of $L$ alone, roughly.  We also find that
$(T_c - T_r)/(T_c - T_n)$ is always in the range $[0.6,0.8]$, so that
reheating to $T_c$ is never achieved.  One reason to be interested
in this is that complete reheating to $T_c$ tends to slow the growth
of the bubbles significantly, and most baryogenesis mechanisms predict
that the baryon asymmetry is enhanced by $1/v$ if the bubble wall
velocity $v$ is small.

\section{Will Electroweak Baryogenesis be Ruled Out (or In) by LEP?}

The most pressing question confronting electroweak baryogenesis is
whether it is really testable at LEP.  Fig.~12 shows the regions in the
$\tan\beta$-$m_{h}$ plane which will be excluded in runs near 200
GeV center of mass energy.  At first sight the experimentally
inaccessible region $m_h > 95$ GeV, $\tan\beta > 10$ might look
worrisome, since these values appear to be allowed by our Monte Carlo.
Closer examination shows that these points only escape detection by LEP
if $m_{A^0}$ and $\sin^2(\alpha-\beta)$ are too small to be compatible
with a strong phase transition.  The real worry is whether $m_h > 107$
GeV, which is above the discovery potential of LEP 200, but still
compatible with baryogenesis if the left-handed stop is heavy enough.
In this case we will have to hope for the slim possibility of the Higgs
being discovered at Tevatron.

\begin{figure}
\centerline{\psfig{figure=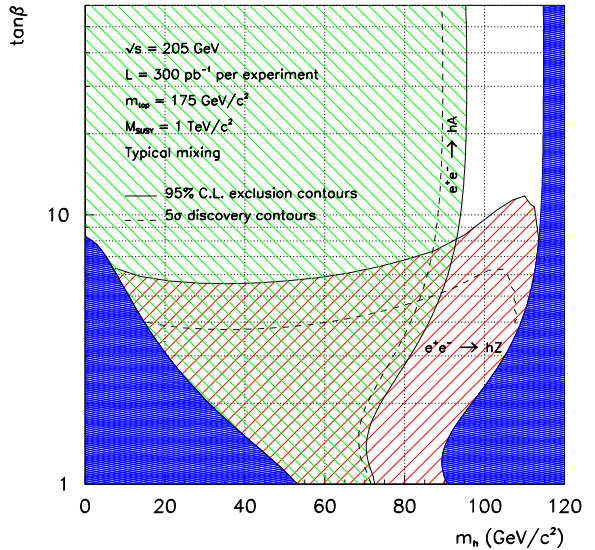,width=3.0in}}
\caption{Discovery potential of LEP 2.}
\end{figure}

\section{Conclusions}

The most promising experimental signal for electroweak baryogenesis is
a light Higgs, with $m_h < 86$ GeV if the left-stop mass parameter
$m_Q$ is 100 GeV, and $m_h < 116$ GeV if $m_Q=2$ TeV.  It is also
possible that the mostly right-handed top squark will be observed,
since its mass is constrained to lie in the range 119 GeV $< m_{\tilde
t} <$ 172 GeV.  This extreme lightness of the stop can only be achieved
by taking its mass parameter $m_U^2$ to be dangerously negative.  The
danger is that the universe will get stuck in a color-breaking minimum
with a nonzero stop condensate, and not be able to tunnel into the
normal vacuum state where we live.  Although it has been previously
investigated,\cite{bjls} this issue probably deserves closer scrutiny.

\section*{References\footnote{see ref.~1 for a more complete list 
of references.}}


\begin{thebibliography}{99}

\bibitem{cm} J.M.~Cline and G.D.~Moore, 
	hep-ph/9806354, to appear in Phys.~Rev.~Lett.\ (1998).
\bibitem{cqw2} M.~Carena, M.~Quiros and C.E.M.~Wagner, hep-ph/9710401,
	Nucl.\ Phys.\ B524, 3 (1998).
\bibitem{cmer} J.M.~Cline, J.R.\ Espinosa, G.D.~Moore and A.\ Riotto,
	preprint hep-ph/9810261 (1998).
\bibitem{jre} J.R.\ Espinosa, , hep-ph/9604320, Nucl.\ Phys.\ B475 (1996) 273 
\bibitem{bjls} D.~B\"odeker, P.~John, M.~Laine and M.G.~Schmidt,
	 hep-ph/9612364, Nucl.~Phys.~B497 (1997) 387.
\bibitem{lr1} M.~Laine and K.~Rummukainen, hep-ph/9804255,
	Phys.\ Rev.\ Lett.\ 80 (1998) 5259 ; hep-lat/9804019,
	preprint CERN-TH-98-122 (1998)
\bibitem{mqs} J.M.\ Moreno, M.\ Quiros and  M.\ Seco, hep-ph/9801272,
	Nucl.\ Phys.\ B526 (1998) 489.
\bibitem{cjk} J.M.\ Cline, M.\ Joyce and K.\ Kainulainen, hep-ph/9708393,
	Phys.\ Lett.\ B417 (1998) 79.

\end{thebibliography}
\end{document}